\documentclass{aastex631}

\shorttitle{Tunguska strewn field}
\shortauthors{Carbognani et al.}

\usepackage{amssymb}
\usepackage{textcomp}
\usepackage{graphicx}
\usepackage{txfonts}
\usepackage{epstopdf} 
\usepackage{xcolor}     % For text colors
\begin{document}

\title{Computation of a possible Tunguska's strewn field}

\author{Albino Carbognani}
\affiliation{INAF - Osservatorio di Astrofisica e Scienza dello Spazio \\
Via Gobetti 93/3 \\
40129 Bologna, Italy}

\author{Mario Di Martino}
\affiliation{INAF-Osservatorio Astrofisico di Torino \\
Via Osservatorio 20 \\
10025 Pino Torinese (TO), Italy}

\author{Giovanna Stirpe}
\affiliation{INAF - Osservatorio di Astrofisica e Scienza dello Spazio \\
Via Gobetti 93/3 \\
40129 Bologna, Italy}

\begin{abstract}
On June 30, 1908, at about 0h 14.5m UTC, the Tunguska Event occurred, most likely caused by the fall of a small stony asteroid of about 50-80 meters in diameter over the basin of the Tunguska River (Central Siberia). The first expedition was made by the geologist Leonid Kulik 19 years after the event, and macroscopic meteorites have never been found around epicenter. In this paper, we want to establish whether stony macroscopic fragments could have survived the TCB's airburst (Tunguska Cosmic Body) and where they might have fallen. For this purpose, we have implemented a fall model to describe the mass ablation, pancake expansion, airburst and fragments's dark flight. In our scenario, the fragments have a higher mean strength than the main body due to Weibull's law. The results, for a TCB with a kinetic energy of 15 Mt, atmospheric entry speed in the range of 11-20 km/s, trajectory inclination of $35^\circ$ and average strength in the range of 3-70 MPa, tell us that for a macroscopic fragment with a mean strength between 14-85 MPa would be possible to survive the high pressure and temperature airburst to reach the ground. The falling speed of the fragments is in the range of 0.8-0.5 km/s, which favoured their burial in the permafrost. The range of mean strength values for the TCB's macroscopic fragment appears physically plausible if we consider the fall of Carancas in 2007, with an estimated strength of 20-40 MPa. So our possible strewn field, computed for a typical fragment's speed of $10 \pm 3 $ km/s, is located about 11 km North-West from the epicenter, with an area of about $140~\textrm{km}^2$. Finally, Cheko Lake, which by some authors is considered an impact crater, falls about 3.5 km outside the strewn fields at 3 sigma level and it is unlikely that it could be a real impact crater: only if the TCB's trajectory had an azimuth in the range $150^\circ - 180^\circ$ would be in the strewn field area, but this is not consistent with the most likely trajectory azimuth.   
\end{abstract}

\keywords{minor planets --- asteroids --- Tunguska}

%%%%%%%%%%%%%%%%%%%%%%%%%%%%%%%%%%%%%%%%%%%%%%%%%%%%%%%%%%%%%%%%%%%%%%%%%%
\section{Introduction}
\label{sec:intro} 

We define as Tunguska-class bodies the near-Earth asteroids (NEAs), with a diameter between 10 and 100 meters, that generally explode in the atmosphere before reaching the ground unless their internal strength is particularly high. In an impact event with our planet, these bodies are not large enough to cause a global climate catastrophe, but thanks to the high fall speed and consequent great kinetic energy dissipation with strong shock waves, they can generate local airbursts that for larger objects in this class, it can be similar or worse than the Tunguska Event (TE) in Central Siberia.\\
In this area, on Jun 30, 1908, at 07:14:28 local time (00:14:28 UT), a small cosmic body of 50-80 meters in diameter exploded into the atmosphere about 8.5 km above the ground. In the explosion, an energy around 15 Mt was developed, which, in the form of a thermal wave and shock wave, destroyed an area of $2150 \pm 50 ~\textrm{km}^2$ of Siberian taiga. The first on-site scientific expedition was conducted in 1927, 19 years after the event, due to the Russian geologist Leonid Kulik \citep{kulik1936}. Despite the careful research conducted around the epicenter, no macroscopic meteorite has ever been found despite the fierce excavation research conducted by Kulik in the area. For more details about TE, in addition to the aforementioned Kulik's paper, see \cite{Zotkin1966}, \cite{Menahem1975}, \cite{Korobeinikov1976}, \cite{Sekanina1983}, \cite{Chyba1993}, \cite{Vasilyev1998}, \cite{Longo2007}, \cite{artemieva2016} and references therein.\\
The most recent was the Chelyabinsk Event (CE) of Feb 15, 2013, the most energetic impact event observed after TE. At 03:20:20.8 UT, a small asteroid with an estimated diameter of about $20 \pm 5 $ m begins the entry into the atmosphere toward the Russian city of Chelyabinsk (Lat. $55^\circ 03^\prime$ N, Long. $61^\circ 08^\prime$ E), located just East of the Ural mountains. About 11 s after entering into the atmosphere, at about 29.7 km altitude and 40 km South of Chelyabinsk, the pressure of the atmospheric shock wave becomes very strong, and the asteroid generated an airburst, fragmenting into several pieces \citep{Popova2013}. We have instrumental data about the CE, e.g. video recordings from surveillance cameras, satellite images and infrasonic data. This is a big difference with respect to the TE: as we will see later, in this case, there are eyewitness reports and seismic and barometric registrations only.\\
Despite the small size of the asteroid causing the CE, a big meteorite weighing about 570 kg was found and collected on the bottom of Lake Chebarkul, about 70 km West of the airburst epicenter. A similar event, with the arrival of macroscopic fragments on the ground, may have happened for the TE also. In this hypothesis, the epicenter area would not be the most suitable place to search for big meteorites. In other words, the devastated area of the Siberian Taiga is not necessarily the right place to look for macroscopic objects. However, until now, macroscopic fragments of the Tunguska Cosmic Body (TCB) have never been found. The best candidate is the so-called John's Stone, a 2.0 m by 1.5 m by 1.0 m in size quartzitic boulder consisting of 98.5\% of SiO2 found by John Anfinogenov on July 19, 1972, near epicenter on Stoykovich Mountain with an estimated mass of about 10~000 kg. Likely, John's Stone is of terrestrial origin \citep{Bonatti2015}, but it cannot be completely ruled out that it is a new type of meteorite \citep{Anfinogenov2014}. Considering the absence of meteorites attributable to the TCB, the hypothesis that it was a comet has several supporters, even if the statistics about the possible heliocentric orbits favour the asteroidal origin \citep{Farinella2001}. \\
However, local eyewitnesses of TE tell a different story: they observed a stone that appeared from ``nowhere'' in the destroyed forest, and several local Evenkis reported about fresh furrows in the epicenter with stones in the furrow walls \citep{Anfinogenov2014}. So, in our opinion, the lack of macroscopic meteorites is not proof of the complete disintegration of the TCB: the time elapsed from the fall to the first Kulik expedition was 19 years, enough time for any little craters and meteorites to be buried by mud and vegetation. Indeed, while numerical simulations of the TE that assume a strengthless falling body can replicate the shape and width of the devastation area by the shock wave \citep{artemieva2007, artemieva2016}, such models cannot preclude the survival of cm-m scale fragments that are too small for the model to resolve. \\
Other authors point out that the dynamic strength of decimetric rock objects entering the atmosphere is of the order of 1 MPa; therefore, the arrival of large rocks on the ground in an energetic event such as that of Tunguska is very unlikely because they would completely disintegrate during the airburst \citep{collins2008}. The issue is controversial, we will discuss it in section \ref{sec:TE}.\\
In this paper, using CE as a guide to test a fall model for macroscopic fragments, we will delimit a possible strewn field to search for possible macroscopic meteorites belonging to TCB. We used the Chelyabinsk event as a test because it is the closest big event to the Tunguska event that we have data on. As far as the authors know, such a computation for Tunguska has never been made because meteorites related to the event have never been found, and therefore, it was assumed that the TCB completely disintegrated, leaving no traces other than high-Z microparticles in the resin of the 1908 trees \citep{Longo1994}. \\
The paper is organized as follows: in Section~\ref{sec:dark}, we will briefly see the physical model adopted for dark flight, strewn field and mass ablation; in Section~\ref{sec:CE}, we will see the CE, a test for the dark flight and impact point model about fragment F1 as well a mean strength estimate for the main body; in Section~\ref{sec:TE} we will see a description of the TE, the determination of the most probable heliocentric orbits and a model for the fall of the TCB which includes mass ablation, pancake phase and airburst. In this section, we will estimate the mean strength a macroscopic fragment must have not to fragment further and reach the ground. In Section~\ref{sec:TE_strewn_field}, we will see the computation of the strewn field for macroscopic fragments and some considerations about the Lake Cheko origin. Finally, we provide our conclusions.

\section{Dark flight, strewn field and ablation model}
\label{sec:dark}
The dark flight phase of a meteoroid falling into the atmosphere begins at the end of the fireball phase when the atmospheric speed drops below about 3 km/s. From elementary physics, the motion equation in a geocentric inertial reference system describing the fall of a meteoroid with mass $m$ and acceleration $\vec{a}_m$ in the dark flight phase is as follows:

\begin{equation}
m\vec{a}_m =  \vec{F}_g + \vec{F}_d = -GMm\frac{\vec{r}}{r^3} -\Gamma\rho_a \left|\vec{V}_m - \vec{W}\right| A\left( \vec{V}_m - \vec{W} \right)
\label{eq:motion_inertial}
\end{equation}

\noindent Eq.~(\ref{eq:motion_inertial}) is a non-linear differential equation in vector form. The first term on the right is the gravity force ($G$ is the gravitational constant, $M$ is the Earth's mass and $\vec{r}$ the distance between the geocenter and the meteoroid), the second is the drag force (Newton's Resistance law) exerted by the air on the meteoroid: $\Gamma$ is the dimensionless drag coefficient (equal to 0.5 for a sphere), $\rho_a$ the air density, $A$ the meteoroid cross section, $\vec{V}_m$ the meteoroid speed and $\vec{W}$ the wind speed. The value of the drag coefficient $\Gamma$ depends both on the unknown final form of the meteoroid after ablation and on the Mach number, i.e., the ratio between the meteoroid speed and the sound speed at the same height above ground \citep{Ceplecha1987}. The value of the drag coefficient is independent of the size, the crucial parameter being the body shape. The $\Gamma$ asymptotic value, i.e., toward very high Mach numbers is left as a free parameter, while for low Mach numbers, i.e. equal or less than 4, we adopt the following Ceplecha's values: $\Gamma(4)=0.58$, $\Gamma(3)=0.62$, $\Gamma(2)=0.63$, $\Gamma(1)=0.50$, $\Gamma(0.8)=0.44$, $\Gamma(0.6)=0.39$, $\Gamma(0.4)=0.35$ and $\Gamma(0.2)=0.33$.\\
Instead of using a reference system with the axes fixed in space, we can use a geocentric reference system but with the axes rotating with the Earth's surface, which is where the observer is located. In this case, two inertial forces must be added: the Coriolis force and the centrifugal force. These two inertial forces are generally negligible with regard to the strewn field location, but we include them for completeness. The motion differential equation in this non-inertial system becomes: 

\begin{equation}
\frac{d\vec{v}_m}{dt} = -GM\frac{\vec{r}}{r^3} -\Gamma\rho_a \left|\vec{V}_m - \vec{W}\right| \frac{A}{m}\left( \vec{V}_m - \vec{W} \right)-2\vec{\omega} \wedge \vec{V}_m - \vec{\omega} \wedge \left(\vec{\omega} \wedge \vec{r}\right)
\label{eq:motion_non_inertial}
\end{equation}

\noindent The first additional term is the Coriolis, and the second is the centrifugal. In Eq.~(\ref{eq:motion_non_inertial}) $\vec{\omega}$ is the vector of the
rotational speed of the Earth. By making explicit Eq.~(\ref{eq:motion_non_inertial}) for a geocentric non-inertial Cartesian reference system with $\hat{z}$ direction along
Earth's rotation axis (so $\vec{\omega}=\omega\hat{z}$ is a vector with only the $\hat{z}$ component), $\hat{x}$ and $\hat{y}$ direction in the Equatorial plane, with $\hat{x}$ towards the Greenwich meridian, six first-order differential equations are obtained, three differential equations for the speed components and three for the position components. 
Solving these equations numerically with a Runge-Kutta 4th/5th order solver will provide the position and velocity of the meteoroid as it fall towards the ground, and the intersection of the trajectory with the ground gives the strewn field position. \\
These are the equations that underlie ``Meteorite Finder'', a software we developed with Matlab 2019b and freely available\footnote{\url{https://github.com/AlbinoCarbo/Meteorite_Finder.git}}, which computes the meteoroid's dark flight and delimits the strewn field on the ground. To work, Meteorite Finder needs an atmospheric profile computed for the starting point of the dark flight with atmospheric pressure, wind $u$ component, wind $v$ component, absolute temperature, relative humidity and height above ground. If desired, different Monte Carlo scenarios with standard normal distribution can be computed to estimate the uncertainty of the strewn field based on the uncertainty of the six starting parameters at the beginning of the dark flight: speed, trajectory inclination above surface and azimuth (counted from north to east), height above Earth's surface, latitude and longitude of the starting point. \\
The dark flight model can easily be extended to describe the fireball phase of a meteoroid by adding to Eq.~(\ref{eq:motion_non_inertial}) the scalar mass loss equation describing the ablation process:

\begin{equation}
\frac{d m}{dt}=\dot{m} = -\Gamma \frac{C_H}{Q} \rho_a A {{V}_m}^3
\label{eq:mass_loss_ablation}
\end{equation}

\noindent In Eq.~(\ref{eq:mass_loss_ablation}) $Q$ is the heat of ablation, $C_H$ is the heat transfer coefficient and the drag coefficient $\Gamma$ in this case is a constant because this equation intervenes at hypersonic speeds only. For stony asteroids $Q\approx 8\cdot 10^6$ J/kg, while $C_H\approx 0.1$ \citep{Chyba1993, Avramenko2014}. If the meteoroid's mass decreases, the value of section $A$ also decreases accordingly. As we said at the beginning, ablation ceases when the velocity drops below approximately 3 km/s. The use of Eq.~(\ref{eq:mass_loss_ablation}) with Eq.~(\ref{eq:motion_non_inertial}) will be useful in the case of TE because the fragment's speed after the airburst phase is high enough to have ablation still.

\section{The Chelyabinsk event}
\label{sec:CE} 
As a first step, we will use the data about the CE to test the dark flight and strewn field computation tools. We aim to predict the impact place of the largest fragment that survived the airbursts to validate the model. The main data available about the CE are listed in Table~\ref{tab:Chelyabinsk_table}. Various individual fragmentations have been recorded between 40 and 30 km of altitude \citep{Borovicka2013}: after the main airburst, around 29.7 km, about 20 fragments emerged from the disruption clouds. The main boulder was destroyed at an altitude of 22 km, while another fragment - F1 in the Borovi\v cka's nomenclature - continued the fall in the densest layers of the atmosphere, survived a maximal dynamic pressure of about 15 MPa at an altitude of about 20 km, and began the dark flight phase at about 12.6 km height with a speed of 3.2 km/s over a point with coordinates $\textrm{Lat}_{\textrm{F1}}=54.9361^\circ$ N and $\textrm{Long}_{\textrm{F1}}=60.5883^\circ$ E \citep{Borovicka2013}. This will be our starting point. The atmospheric path ended on the frozen surface of Lake Chebarkul $-$ with an ice layer thickness of about 70 cm $-$ opening a hole of about 6 m in diameter at coordinates $\textrm{Lat}_{\textrm{hole}}=54.95976^\circ$ N and $\textrm{Long}_{\textrm{hole}}=60.32087^\circ$ E. The azimuth of the trajectory of this fragment is practically identical to that of the progenitor body (Table~\ref{tab:Chelyabinsk_table}), with a deviation of about 1.3 degrees. \\
After the airburst, the major fragments continue to follow about the same trajectory as the original body, but there are several effects which cause a dispersion: gravity and atmospheric drag, lift or bow shock interaction. Drag and gravity forces are dominant for trajectory inclination angle less than about $30^\circ$, while for higher inclinations, the separation is due to the bow shock interaction that produces a transverse acceleration which separates the fragments from each other \citep{Passey1980}. \\
In the CE case, with an inclination of the trajectory well below $30^\circ$, gravity and atmospheric drag forces dominated in separating the fragments. Fragments of about 0.1 g fall near the point of the main explosion, masses of about 100 g fell further along the trajectory, and at least one of 3.4 kg fall near Timiryazevskiy \citep{Popova2013}. All these minor fragments were subject to acceleration along the trajectory given by Eq.~(\ref{eq:motion_non_inertial}), and considering that the mass-area ratio $m/A$ is proportional to the fragment's radius, the smaller the meteoroid and greater the negative drag acceleration will be. This caused smaller fragments to fall much earlier than the fragment F1.
 
\begin{figure}[ht!]
\centering
\hspace*{-1.6cm}\includegraphics[width=1.2\hsize]{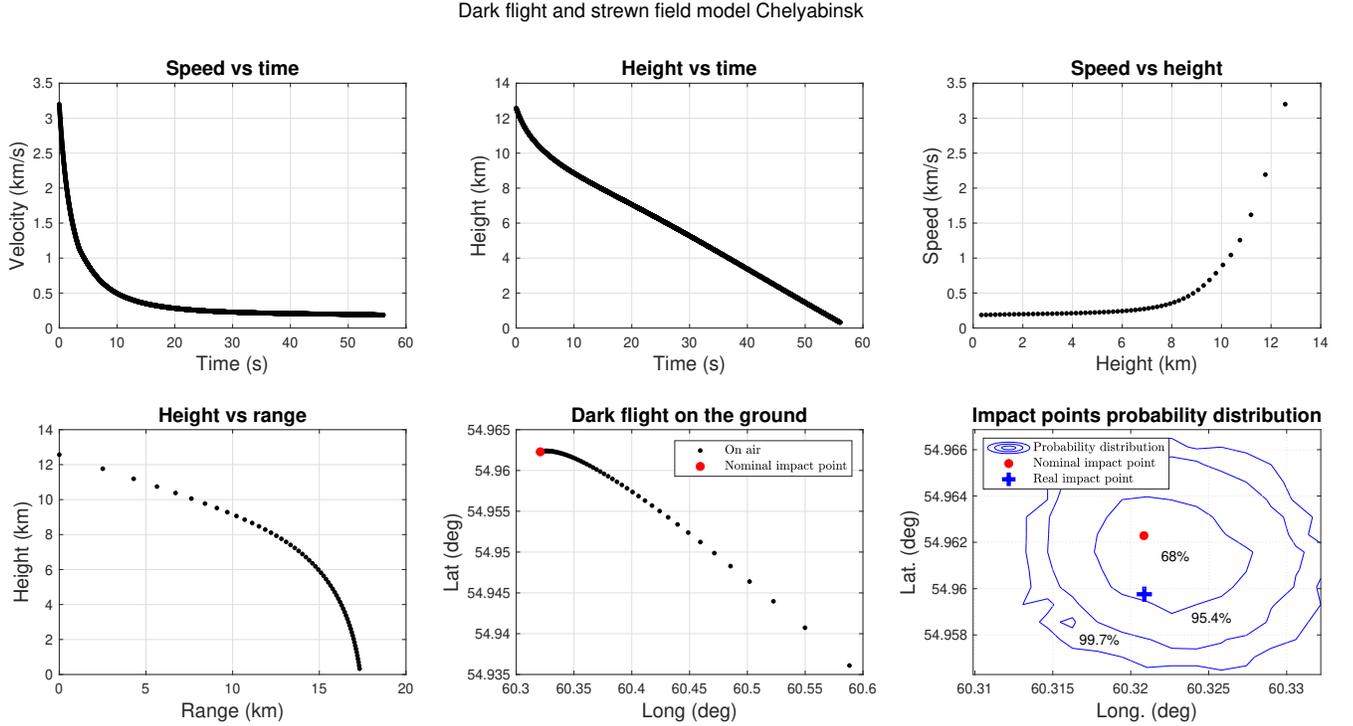}
\caption{The dark flight model for the F1 fragment with $m \approx 650 ~\textrm{kg}$ and $m/A \approx 1580 ~\textrm{kg}/\textrm{m}^2$. The computed nominal impact point, with a number of Monte Carlo scenarios equal to 5000, is at coordinates Lat. $54.962 \pm 0.002$, Long. $60.321 \pm   0.003 $. From the probability distribution of the impact points (bottom right box), we see that the nominal impact point is in the area within which 68\% of the virtual impact points fall, so we can say that it falls approximately within 1 $\sigma$ from the observed impact point.}
\label{fig:Traj_Chel}
\end{figure}

\subsection{Dark flight of the F1 fragment}
The fragment F1 from the CE was recovered in Lake Chebarkul on Oct 16th and found to weigh about 570 kg. Considering that the bulk density of the Chelyabinsk meteorites is $3290 ~\textrm{kg}/\textrm{m}^3$ \citep{Kohout2013}, an effective size of about 0.7 m and a mass-area ratio of about $1500 ~\textrm{kg}/\textrm{m}^2$ can be estimated. Taking into account the residual fragmentation during impact, the impacting mass may have been about 650 kg \citep{Popova2013}, for an original mass-area ratio of about $1580 ~\textrm{kg}/\textrm{m}^2$.\\
We will describe the three-dimensional motion of the main fragment F1 using Eq.~(\ref{eq:motion_non_inertial}). As altitude, inclination, azimuth and starting speed for the dark flight computation, we use the data reported in Table~\ref{tab:Chelyabinsk_table} for F1. To estimate the uncertainty associated with the theoretical impact point, we used a Monte Carlo technique with a standard normal distribution of the parameters altitude, inclination, azimuth and starting speed and with the uncertainties reported in Table~\ref{tab:Chelyabinsk_table}. We computed the dark flight using 5000 different scenarios. In order to model the dark flight phase, it is important to know the atmospheric profile in the data, time and place closest to the fall because the residual meteoroid trajectory can be heavily influenced by the atmospheric conditions, especially in the case of strong winds in the lower layers of the atmosphere. The influence of weather conditions diminishes as the size and speed of the fragment increases. \\
The wind profile for the CE was taken from atmospheric sounding conducted by the Verkhneye Dubrovo station ($56.73^\circ$ N, $61.06^\circ$ E) $-$ the weather station nearest to Chelyabinsk $-$ at 0:00 UT. With these input data, the computed impact coordinates for F1 are long. $60.321^\circ\pm 0.003^\circ$ N and lat. $54.962^\circ\pm 0.002^\circ$ E with the asymptotic value of drag coefficient equal to $\Gamma\approx 0.775$. These impact coordinates are about 17.2 km away from the start of the dark flight and about 300 m from the observed impact point, i.e. the computed impact point is within 1 sigma from the observed one, see Fig.~\ref{fig:Traj_Chel} for the results about the dark flight model. Considering that the weather profile used was quite far from the starting point and that the inclination of the trajectory was very low, this is a good result. We will apply the same technique, adding to Eq.~(\ref{eq:motion_non_inertial}) also the Eq.~(\ref{eq:mass_loss_ablation}), to estimate the position and extent of the possible Tunguska strewn field.

\begin{table}
\centering
\caption{Data about the atmospheric trajectory of the Chelyabinsk event and the hole in the Chebarkul lake \citep{Popova2013}. The data about the major fragment F1 refer to the beginning of the dark flight phase \citep{Borovicka2013}. Azimuth is clockwise from North.}
\label{tab:Chelyabinsk_table}

\begin{tabular}{lc}
\hline
Quantity & Best Value \\
\hline
$H_{start}$ (km)                        & $97.1 \pm 2$ \\
$v_\infty$ (km/s)                       & $19.2 \pm 0.3$ \\
Entry elevation angle ($^\circ$)        & $18.3 \pm 0.2$ \\
Entry azimuth ($^\circ$)                & $103.2 \pm 0.4$\\
Entry latitude ($^\circ$)               & $54.44 \pm 0.02$ N\\
Entry longitude ($^\circ$)              & $64.56 \pm 0.03 $ E\\
\hline
Main peak altitude (km)                 & $29.7 \pm 0.7$\\
Main peak latitude ($^\circ$)           & $54.84 \pm 0.02$ N\\
Main peak longitude ($^\circ$)          & $61.41 \pm 0.03$ E\\
Main peak speed (km/s)                  & $19.2 \pm 0.2$ \\
\hline
F1 height (km)                          & 12.57\\
F1 speed (km/s)                         & 3.2 \\
F1 latitude ($^\circ$)                  & 54.9361 \\
F1 longitude ($^\circ$)                 & 60.5883 \\
F1 azimuth ($^\circ$)                   & $101.87 \pm 0.4$ \\
F1 elevation angle ($^\circ$)           & $17.53 \pm 0.3$ \\
\hline
Hole Chebarkul Lake latitude ($^\circ$) & $54.95976 \pm 0.00006$ N\\ 
Hole Chebarkul Lake longitude ($^\circ$)& $60.32087 \pm 0.00006$ E\\
Hole diameter (m)                       & $6.3 \pm 0.5$ \\
\hline	
\end{tabular}
\end{table}

\subsection{Strength of the Chelyabinsk event}
Usually, the meteoroids fragmentation model assumes that the process starts when the aerodynamic pressure from Eq.~(\ref{eq:motion_inertial}) $P_{dyn}=\Gamma \rho_{fr} {V_m}^2$ in front of the body is equal or superior to mechanical strength $S$ of the body: $P_{dyn}\geq S$ \citep{Passey1980, Chyba1993, Hills1993, Svetsov1995, Farinella2001}. As before, the quantity $\rho_{fr}$ is the air density at the fragmentation level, $V_m$ is the body's speed with respect to air and $\Gamma$ is the dimensionless drag coefficient. So the starting fragmentation condition is: 

\begin{equation}
S = \Gamma\rho_{fr}{V_m}^2  
\label{eq:strength}
\end{equation}

\noindent In CE, we can estimate $S$ using the strength measured on the recovered meteorites, but it must be kept in mind that the strength of the meteorites does not coincide with the body's mean strength. For example, in the case of the major fireballs observed by the European Fireball Network, the original meteoroid has a dynamic strength always lower with respect to the subsequent fragmentations: the strength of the first fragmentation is in the range 0.4-4 MPa while the main fragmentation is in the range 3.5-12 MPa \citep{Borovicka2008}. This difference is probably due to fractures present in large meteoroids or small asteroids, which decrease the overall strength of the original body compared to the individual non-fractured blocks \citep{Popova2011}. Put in a different way, assuming that the structure is monolithic and not a rubble pile, the mean strength of the original body is the lower limit of the mean strength of the blocks into which it can separate. Considering that the cohesionless spin-barrier of about 2.2 h exists only for asteroids with a diameter greater than about 150 m \citep{PravecHarris2000}, it is generally believed that small asteroids are monolithic type, so this category includes the asteroids responsible for the Chelyabinsk and Tunguska events. It should be noted that this distinction between monolithic asteroids and rubble piles, depending on diameter, is not so categorical. Molecular cohesion forces, mediated by dust grains, can also intervene to keep the blocks of a rubble pile asteroid together and make them violate the spin barrier \citep{Scheeres2010}. In our case, we assume that the TCB was monolithic.\\
If the mechanical destruction of a monolithic body occurs along the fracture lines, then it is reasonable to expect that the strength depends on the volume or mass and that the following scale relation holds \citep{Weibull1951, Svetsov1995, Scheeres2015, Jenniskens2022}:

\begin{equation}
S_{main} = S_{fr}\left( \frac{m_{fr}}{m_{main}} \right)^{\alpha}
\label{eq:scale_strength}
\end{equation}

\noindent This law of scaling, based on Weibull's statistic, derives from the fact that if the mass decreases, the size of the weak points in the rock must also decrease, so the strength must increase. In Eq.~(\ref{eq:scale_strength}) $\alpha$ is the Weibull modulus, $S_{main}$ is the main body strength while $S_{fr}$ is the fragment strength. The $\alpha$ value increases as the inhomogeneity of the material increases, and a value between 0.1 and 0.7 is expected. \\
In the case of Chelyabinsk, the $\alpha$ value can be directly estimated from the mechanical strength and mass values measured on meteorites. Mechanical properties of the Chelyabinsk meteorite were determined at NASA Ames, and compression strength is determined to be 330 MPa (C3: 4.46 g), 327 MPa (C4: 4.84 g) and 408 MPa (C5: 1.58 g), similar to other ordinary chondrites \citep{Popova2013}. Using Eq. (\ref{eq:scale_strength}) and these meteorites data we get $\alpha\approx 0.2$, a reasonable value within the expected range. \cite{Avramenko2014}, modeling the fall and fragmentation of the CE, finds $\alpha\approx 0.18$, in excellent agreement with our estimate. For comparison, in the case of the Almahata Sitta meteorites, $\alpha = 0.3$ is found for the asteroid 2008 TC3 \citep{Jenniskens2022}. So from the total mass estimated for Chelyabinsk, about $1.4\cdot 10^7$ kg and from the average strength of the two major samples, we find $S_{CE}\approx 4$ MPa as in \cite{Foschini2019}, which uses the same method. This value is of the same order of magnitude as that found by \cite{Borovicka2013} with $S\approx 1$ MPa.

\section{The Tunguska event}
\label{sec:TE}
As we have already mentioned in the introduction, while for the CE, the available data are abundant, for the TE instrumental values are scarce anyway, the most reasonable data are reported in Table~\ref{tab:Tunguska_table}. The uncertainties reported in this table $-$ apart the speed values $-$ are the standard deviation of the mean values given by various authors and summarized in table 18.1 of the \cite{Longo2007} work. Our paper is not intended to re-discuss the results of Tunguska, so we limit ourselves to some essential points to understand the starting parameters that we can consider reasonably sure.\\ 
About the TE, there are seismic and barometric registrations recorded immediately after the event and data on forest devastation about directions of flattened trees and charred trees collected in a century of expeditions. The time of the event was established reasonably well from the seismic and barometric recordings. The most widely quoted magnitude range of the developed energy, based on historic barograms, seismic records, and forest damage compared with nuclear airbursts, is between 10 and 40 Mt with a most probable value of about 15 Mt \citep{Vasilyev1998}, even though \cite{Boslough1997, Boslough2008}, on the basis of the results of numerical simulations, estimated an energy of $3-5$ Mt. From forest devastation and different arrival times for Rayleigh and SH body waves recorded at Irkutsk, an explosion height of about 8.5 km was obtained \citep{Menahem1975}. The geographic coordinates on the ground of the explosion in the atmosphere (the so-called epicenter) were set by the azimuth distribution of the flattened trees, while from the symmetry of the devastation area and eyewitness data, the range for trajectory azimuth and inclination above Earth surface was set \citep{Farinella2001, Longo2007}. \\
The uncertainty interval of the arrival azimuth is relatively small: roughly, the body arrived along a trajectory from south-east to north-west. On the other hand, it is much more complex to establish the trajectory inclination. From the analysis of the eyewitnesses collected mostly about 50 years after the event, low-inclination trajectories in the range $h_i = 5^\circ - 17^\circ$ are found. Eyewitness accounts quickly become unreliable, so they should be cautiously treated. In fact, low inclination trajectories do not allow recovery of the disposition observed for the fallen trees. In 1966 \cite{Zotkin1966} tried to reproduce the fallen trees azimuths with experiments on a model forest simulated in the laboratory by superimposing a spherically symmetrical shock wave of the airburst with the cylindrical shock wave coming from the arrival trajectory. The figure of the fallen trees had the greatest resemblance to that observed for inclinations around $30^\circ$. Also, the numerical simulations on the fallen trees conducted by \cite{Korobeinikov1976} provided a preferential value of the inclination around $40^\circ$ and in their simulations \cite{artemieva2007, artemieva2016} assume inclinations of $30^\circ$ and $45^\circ$. Therefore, the value of the trajectory inclination is much more uncertain than the azimuth, and it is for this reason that in Table~\ref{tab:Tunguska_table} there are two different values. In the computations concerning the dynamic pressure and the strewn field, we will favour the high inclination trajectories.\\

\begin{table}
\centering
\caption{Data about the TE. The uncertainties reported in this table $-$ apart the speed values $-$ are the standard deviation of the mean values given by various authors and summarized in table 18.1 of the \cite{Longo2007} work. The RA and DEC of the apparent radiant have been computed starting from the trajectory azimuth and inclination from the epicenter point. The epicenter coordinates came from the data about fallen trees \citep{Fast1967, Longo2005}.}
\label{tab:Tunguska_table}

\begin{tabular}{lc}
\hline
Quantity & Best Value \\
\hline
Entry inclination angle, $h$ ($^\circ$)    & $15 \pm 5$ (eyewitnesses); $35 \pm 5$ (simulation)\\
Entry azimuth$^{a}$ ($^\circ$)        & $110 \pm 10$\\
RA apparent radiant ($^\circ$)        & $89 \pm 10$ (eyewitnesses); $80 \pm 10$ (simulation)\\
DEC apparent radiant ($^\circ$)       & $4 \pm 5$ (eyewitnesses); $21 \pm 5$ (simulation)\\
\hline
Airburst time (UT)                & $0$h $14$m $28$s\\
Airburst height (km)              & $8.5 \pm 1$\\
Airburst energy (Mt)              & $10-40$; $\sim 15$\\
Epicenter Latitude ($^\circ$)         & $60.886 \pm 0.002$ N\\  % Fast epicenter: 60° 53' 09" N confermato da rianalisi di Longo et al 2011
Epicenter Longitude ($^\circ$)        & $101.894 \pm 0.002$ E\\ % Fast epicenter: 101° 53' 40" E
Most probable speed range (km/s)      & $11-20 $ \\
\hline	
\multicolumn{2}{l}{$^{a}$ Clockwise from North}\\
\end{tabular}
\end{table}

\subsection{TCB's atmospheric entry speeds and orbits}
\label{sec:TE_orb}
The most critical parameter to set, very important both to obtain the heliocentric orbit and the dark flight trajectory of hypothetical macroscopic fragments, is the TCB's atmospheric entry speed that, if the body belongs to the Solar System, can range from about 11.2 to 72 km/s \citep{Passey1980}. Taking into account the range parameters of the trajectory (azimuth, inclination, airburst height) given in Table~\ref{tab:Tunguska_table}, we used a Monte Carlo technique with standard normal distribution of the parameters to explore five different atmospheric entry speed ranges, $V_m = 16 \pm 5$, $20 \pm 5$, $30 \pm 5$, $35 \pm 5$ and $40 \pm 5$ km/s, to compute the possible heliocentric orbits as in \cite{Carbognani2021}. For higher atmospheric entry speeds, the TCB no longer belongs to the Solar System because the heliocentric speed is higher than the escape velocity from the Sun at the Earth's distance (about 42 km/s). Assumed a trajectory inclination (low $15^\circ \pm 5^\circ$ or high $35^\circ \pm 5^\circ$), for each speed interval, 5000 clones with the parameters compatible with the observed ground trajectory have been extracted, and correcting it for Earth's rotation and gravity attraction \citep{Ceplecha1987}, the corresponding 25,000 heliocentric orbits have been computed. Considering only closed orbits, they were classified using the Tisserand parameter with respect to Jupiter ($T_J$) because, in first approximation, it is able to discriminate between the asteroidal orbit ($T_J \geq 3$) and cometary orbit ($T_J < 3$). The results are shown in Table~\ref{tab:Tunguska_orbits}. \\
In the case of low inclination trajectory ($h_i =15^\circ \pm 5^\circ$), about 60\% of the 18,055 closed orbits are of asteroidal origin, while the remaining 40\% are of cometary origin. For values of the atmospheric entry speed around 40 km/s, the closed orbits are about 30\%, i.e. the TCB becomes mostly an interstellar object. For high inclination trajectory ($h_i =35^\circ \pm 5^\circ$), the closed orbits are 21,405, and the overall percentage of asteroid orbits rises to 72\%, while cometary orbits are about 28 percent. In this case, for the TCB, it is more difficult to have an open orbit because, for speeds around 40 km/s, the closed orbits are still 55 percent. Our results for high inclination are in good agreement with \cite{Farinella2001} that exploring the ranges $h_i = 3^\circ-5^\circ$, $V = 14-16$ km/s and $h_i = 15^\circ-28^\circ$, $V = 30-32$ km/s and considering $T_J$ as a parameter for closed orbits classification had found 77\% of asteroidal orbits and 23\% of cometary orbits. Note that in Farinella's paper, the high inclination range is greater than the low inclination range, so the high inclination orbits dominate, and that is why Farinella's results agree better with our results for high inclinations. From a statistical point of view, the asteroidal origin for the TCB is the most favoured one. \\

\begin{table}
\centering
\caption{The percentage of asteroid or comet type heliocentric orbits as a function of the TCB's atmospheric entry speed. Considering the closed orbits, for low inclination trajectory, about 60\% are asteroid-type, while the remaining 40\% are comet-type. For high inclination trajectory, the percentages become 72\% and 28 percent.}
\label{tab:Tunguska_orbits}

\begin{tabular}{lccc}
\hline
Trajectory inc. $15^\circ \pm 5^\circ$ & & & \\
\hline
$V~(\textrm{km}/\textrm{s})$ & Asteroids (\%) & Comets (\%) & Closed orbits (\%)\\
\hline
$16 \pm 5$    & 96.6 & 03.4  & 99.8 \\
$20 \pm 5$    & 90.6 & 09.4  & 98.7 \\
$30 \pm 5$    & 55.4 & 44.5  & 77.2 \\
$35 \pm 5$    & 36.9 & 63.1  & 55.6 \\
$40 \pm 5$    & 21.5 & 78.5  & 29.8 \\
\hline	
\hline
Trajectory inc. $35^\circ \pm 5^\circ$ & & & \\
\hline
$V~(\textrm{km}/\textrm{s})$ & Asteroids (\%) & Comets (\%) & Closed orbits (\%)\\
\hline
$16 \pm 5$    & 99.4 & 0.6  & 100 \\
$20 \pm 5$    & 97.6 & 2.4  & 99.8 \\
$30 \pm 5$    & 75.4 & 24.6 & 93.5 \\
$35 \pm 5$    & 54.9 & 45.1 & 79.4 \\
$40 \pm 5$    & 34.9 & 65.1 & 55.4 \\
\hline
\end{tabular}
\end{table}

\subsection{TCB's fall models}
\label{sec:TE_models}
The first to consider the TCB as an asteroidal body with high internal cohesion was Zdenek Sekanina \citep{Sekanina1983}. Sekanina, after having rejected the hypothesis of a cometary TCB because it would have been too fragile to reach the troposphere, considers a stony asteroid of 90-190 meters in diameter with an atmospheric entry speed of about 10 km/s.\\
In 1993, Christopher Chyba and colleagues resumed the idea of an asteroidal TCB characterized by a constant internal strength $S$ described using a ``pancake model'', so called because the impactor, from a certain point onwards, enlarges its radius during the fall \citep{Chyba1993}. The TCB loses its kinetic energy for two reasons: a) the decrease in speed due to drag b) the ablation of the asteroid surface due to the hot frontal shock wave, two physical processes described by Eq.~(\ref{eq:motion_inertial}) and Eq.~(\ref{eq:mass_loss_ablation}). For stony objects in the range 10-100 m in diameter, when the pressure exerted by the shock wave exceeds the strength of the body, first deformation and then atmospheric fragmentation occurs: the hemisphere of the body in direct contact with the shock wave is subject to a dynamic pressure given by $P_{dyn} \approx \Gamma \rho_a {V_m}^2$, while the opposite hemisphere feels a much lower pressure. So it is for this pressure difference that the object is ``crushed and deformed'', becoming a pancake. For higher dimensions, the pressure wave induced on the asteroid does not have time to cross it all before it reaches the ground: in this case, there is the formation of an impact crater, not an airburst \citep{Chyba1993}. \\
In Chyba's model, to describe the deformation of the meteoroid, a cylinder with a diameter equal to the height is used which, when the dynamic pressure is half of the limit given by Eq.~(\ref{eq:strength}), begins to deform and expand as a plastically deformable body, increasing the area $A$ which intercepts the atmosphere. So after the fragmentation, the body's mass spreads over a greater area, the quantity of intercepted atmosphere increases and therefore, the braking and ablation increases: the body loses kinetic energy very rapidly, i.e. explosively, and there is an airburst. The airburst occurs when, at the end of the pancake phase, the fragmented asteroid has a diameter 5-10 times that of the original body, and the fragments decouple, developing their own shock wave. According to Chyba's estimates, for a rocky asteroid with a $45^\circ$ inclined trajectory, 15 km/s speed, 10 MPa strength and starting kinetic energy of 15 Mt, the airburst occurs around 9 km altitude, in good agreement with what was determined for TE. \\
It should be noted that in the pancake model, the mean body strength is much lower than the value that would be obtained by applying Eq.~(\ref{eq:strength}) at the airburst height. This happens because the height where the fragmentation of the main body begins is higher than the height where the maximum expansion of the pancake, associated with the airburst, occurs. For example, considering that an altitude of 9 km corresponds to an average air density of $0.467 ~\textrm{kg}/\textrm{m}^3$, the ``effective strength'' of the TCB would be about 50 MPa, 5 times greater than the 10 MPa assumed for the mean strength.\\
In Chyba's model, for a carbonaceous asteroid with the same starting condition but with 1 MPa strength, the altitude of the explosion would be around 15 km, while a metallic asteroid with an assumed strength of 100 Mpa, would reach the ground, forming an impact crater. In Chyba's pancake model, the strength of the asteroid is assumed constant, Weibull's law given by Eq.~(\ref{eq:scale_strength}) is not considered, and nothing is said about the possible fragments capable of reaching the ground and becoming meteorites.\\ 
With a similar model \cite{Hills1993} examine the fragmentation of small asteroids in the atmosphere considering metallic bodies ($S\approx 200$ MPa), hard stones ($S = 10-50 $ MPa) and soft stones ($S\approx 1$ MPa), but always with a constant strength $S$. In this model the fragmentation condition is given by $S = \rho_{fr}{V_m}^2 $, i.e. Eq.~(\ref{eq:strength}) without including $\Gamma$ (or with $\Gamma=1$), and once it starts it is a continuous process and ends only when the speed drops below the critical value $V_{crit}=\sqrt{S/\rho_{a}(0)} $, where $\rho_{a}(0)\approx 1.22  ~\textrm{kg}/\textrm{m}^3$ is the air density at sea level. Also, in this case, the strength $S$ is a constant value, which does not increase as the size of the fragments decreases. From dimensional considerations, the post-breakup dispersal speed, $V_{disp}$, of the meteoroid fragments in the pancake phase is given by:

\begin{equation}
V_{disp} \approx \sqrt{\frac{7}{2}\frac{\rho_a}{\rho_m}}V_m
\label{eq:dispersal_speed}
\end{equation}

\noindent where $\rho_m$ is the meteoroid mean density. Fixed the value of $V_m$, the lateral speed increases as the air density increases, therefore, if the fragmentation occurs at low altitude, the lateral expansion speed is greater. The diameter of the pancake structure from the break time $t=0$ is:

\begin{equation}
D \approx D_0 + 2\int_{0}^{t} V_{disp}dt
\label{eq:pancake_diameter}
\end{equation}

\noindent This diameter growth over time increases the drag of the fragmented asteroid and causes a sudden decrease in speed. When the diameter of the pancake structure becomes larger than a certain number of times the original diameter, it is assumed that the fragments decouple and that each fragment continues to fall as an independent body with its own shock wave.  \\
In the model of \cite{Hills1993}, the dimensions of the meteorites on the ground are also estimated, and for the TCB (hard stone, speed 15 km/s, diameter 80 m), it is found that, at most, there can be meteorites with a mass of the order of 1 kg. This is a direct consequence of assuming a constant strength in the model: the body continues to fragment because the $S$ value does not rise after each fragmentation. It is interesting to note that for the Chelyabinsk event (soft stone, speed 20 km/s, diameter 20 m), the \cite{Hills1993} model would predict meteorites with the largest mass of the order of 0.001 kg, a value that is not consistent with the mass of the big F1 fragment.\\
More recently, the description of the Tunguska fall was also made using three-dimensional hydrodynamic simulations, which consider the asteroid as a body without internal cohesion \citep{artemieva2007, artemieva2016}. Compared to the single-body models, the hydrodynamic has the advantage of describing the propagation of atmospheric shock waves created by hypersonic flight and the interaction of shock waves with the surface; therefore, it can potentially reproduce the devastation of the area around the epicenter. Two different conditions have been simulated: an impact with an inclination of $45^\circ$ and a speed of 20 km/s for a body of chondritic composition with a diameter of 50 m (total energy of 10 Mt) and a second scenario with an inclination of $30^\circ$, a speed of 15 km/s and diameter of 80 m (energy 20 Mt). Based on this, the TCB began to deform due to the dynamic pressure at an altitude of about 35 km, at 20 km, it assumed a ``pancake'' shape, and at 15 km, it transformed into a gaseous mixture of dust and air at high temperature (about 10~000 K) that froze at about 6-8 km altitude, in good agreement with the altitude of the Tunguska's airburst. The maximum pressure on the ground exceeds the standard atmospheric pressure by 50\% within an area of 5 km in diameter, therefore, the pressure exerted by the arrival of the shock wave in the epicenter can be estimated as about 50 kPa. As we have said, the limit of a hydrodynamic model is that it does not take into account the cohesion force of the asteroid, as these models treat the meteoroid as a continuum \citep{artemieva2016}. So, the absence of macroscopic meteorites from this model does not mean that the body was completely destroyed.\\
Suppose we are in the airburst phase, with the asteroid completely disintegrated into several fragments, which are immersed in a bubble of hot plasma due to the sudden conversion of kinetic energy into heat. We can estimate the ablation survival time of a TCB's meteoroid of diameter $D$ and mean density $\rho_m$ immersed in this high-temperature environment as described by the previous hydrodynamic model. From energy conservation the maximum ablation rate $\dot{m}$ will be given by \citep{Chyba1993}:

\begin{equation}
\dot{m} = \frac{A_s\sigma T^4}{Q}
\label{eq:ablation_max}
\end{equation}

\noindent In Eq.~(\ref{eq:ablation_max}) $A_s=4\pi (D/2)^2$ is the surface area, $\sigma = 5.67 \cdot 10^{-8} ~\textrm{W}/\textrm{m}^2\textrm{K}^2$ the Stephan-Boltzmann constant, $T$ the absolute temperature of the surface assumed equal to that of the surrounding environment and $Q\approx 8\cdot 10^6 ~\textrm{J}/\textrm{kg}$ is the heat of ablation for stony meteorites. The minimal survival time will be:

\begin{equation}
\Delta t_{min} \approx \frac{m}{\dot{m}} = \frac{\rho_m D Q}{6\sigma T^4}
\label{eq:ablation_time}
\end{equation} 

\noindent With $D \approx 1$ m, $\rho_m\approx 3300 ~\textrm{kg}/\textrm{m}^3$ and $T\approx 10^4$ K from Eq.~(\ref{eq:ablation_max}) we get $\dot{m}\approx 50$ kg/s and from Eq.~(\ref{eq:ablation_time}) the minimal survival time is $\Delta t_{min}\approx 8$ s, a time longer than it takes the fragments to exit from the pancake phase which, for TE, has a duration of about 1 s, as we will see in Section~\ref{sec:maximum dynamic pressure}. So big macroscopic fragments, of at least 0.3 m in diameter, can survive the airburst, provided that the strength of the stones is greater than the value given by Eq.~(\ref{eq:strength}), otherwise the stones will undergo further fragmentation and small fragments can be vaporized.\\
In this regard, an interesting paper is that of \cite{Svetsov1998} concerning the survival of the fragments of the TCB after the airburst. Svetsov has taken a TCB with a diameter of 58 m, speed of 15 km/s, trajectory inclination $45^\circ$ and a mean density of $3500 ~\textrm{kg}/\textrm{m}^3$. Considering the radiation emitted in the airburst and absorbed by the fragments, the conclusion is that fragments up to 10 cm in diameter are vaporized in the airburst, while larger ones can survive. This result is in good agreement with our raw estimates. Based on Eq.~(\ref{eq:ablation_max}), a body with about 10 cm in diameter has a mass of almost 2 kg and an estimated lifetime of $2/50 \approx 0.04 s$, a time much shorter than the pancake phase of about 1 s. In our model, we assumed a diameter of approximately 1.4 m for the initial size of the macroscopic fragment, an intermediate value between the Chelyabinsk F1 fragment and the Carancas meteorite \citep{Borovicka2008}. As we will see in Section~\ref{sec:maximum dynamic pressure}, in the case of Chelyabinsk, a fragment of this starting size, after ablation, becomes very similar to F1.

\begin{figure}[ht!]
\centering
\hspace*{-1.6cm}\includegraphics[width=1.2\hsize]{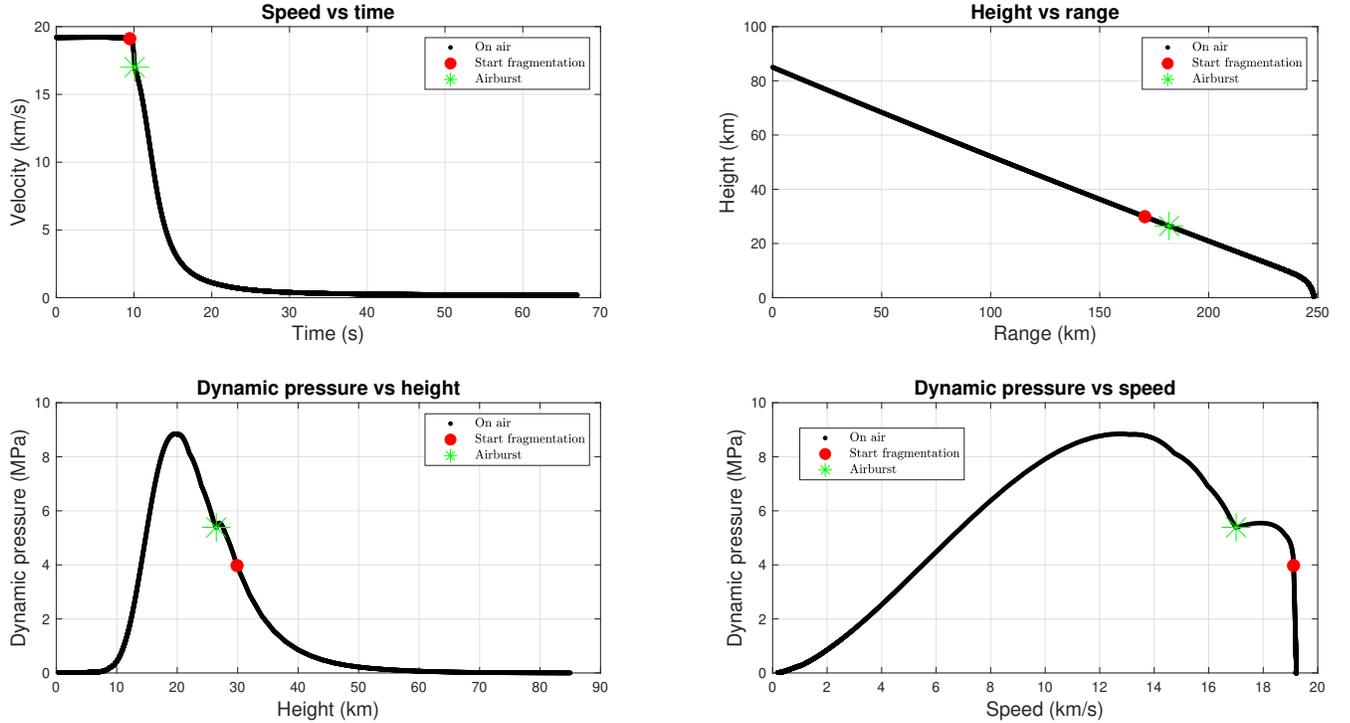}
\caption{The Chelyabinsk model with initial height 85 km, speed 19.2 km/s, $\Gamma = 0.58$, mean density $3290 ~\textrm{kg}/\textrm{m}^3$, diameter 20 m, inclination $18.3^\circ$, pancake factor 7.5 and mean strength 4 MPa. Airburst at 26.5 km, fragment of 1.4 m initial diameter with strength of 19 MPa and final diameter of 0.8 m.}
\label{fig:CE_model}
\end{figure}

\subsection{TCB's maximum dynamic pressure}
\label{sec:maximum dynamic pressure}
\begin{figure}[ht!]
\centering
\hspace*{-1.6cm}\includegraphics[width=1.2\hsize]{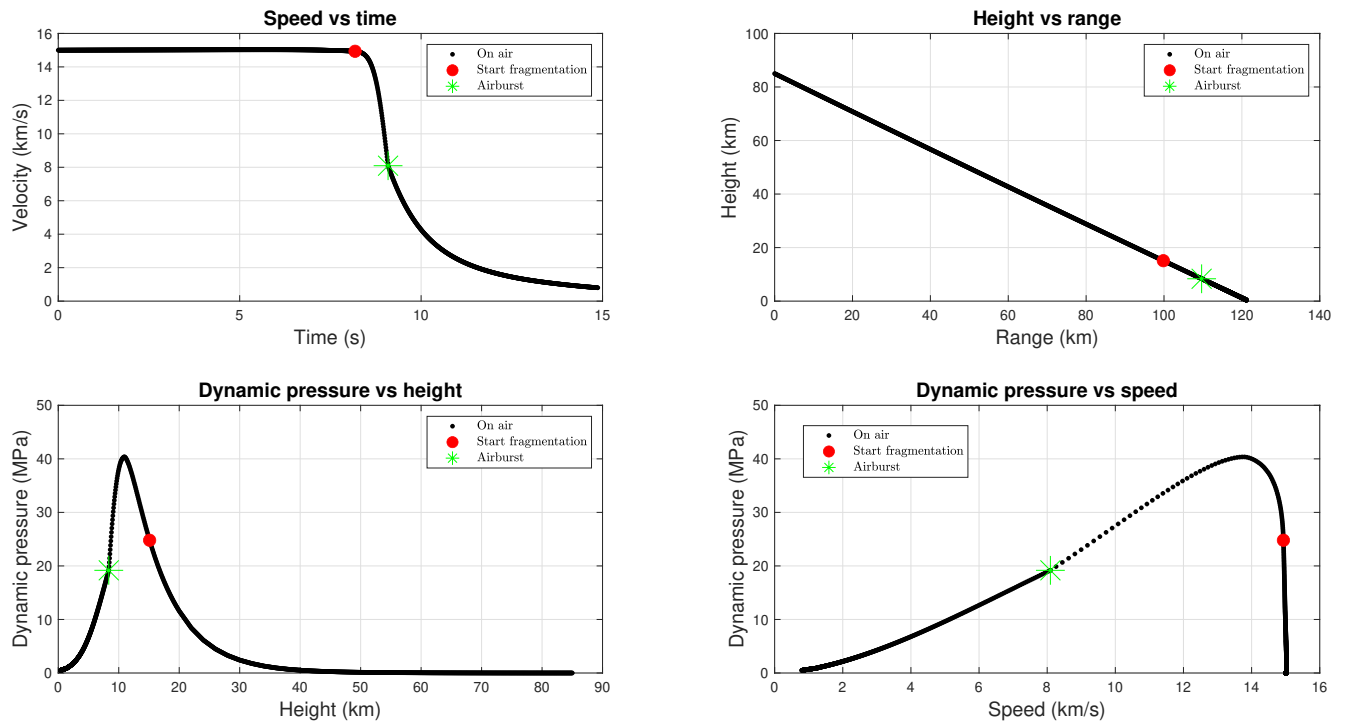}
\caption{The Tunguska model with initial kinetic energy 15 Mt, height 85 km, speed 15 km/s, $\Gamma = 0.58$, mean density $3290 ~\textrm{kg}/\textrm{m}^3$, diameter 69 m, inclination $35^\circ$, pancake factor 7.5 and TCB's mean strength 25 MPa. Duration of the pancake phase, from the fragmentation to the airburst, 0.9 s. Airburst at 8.3 km and maximum dynamic pressure 40.4 MPa. A fragment of 1.4 m diameter with a strength of 100 MPa touches the ground with a diameter of about 1.2 m at a speed of about 0.7 km/s.}
\label{fig:TE_model_15}
\end{figure}

In this section, we want to estimate the dynamic pressure to which a macroscopic fragment from the main body could have been subjected after the fragmentation until the airburst. As stated in section \ref{sec:TE_orb}, considering the range of the possible atmospheric trajectories, the TCB's speed cannot be greater than 35-40 km/s because otherwise, the body could not belong to the Solar System. Assuming that TCB was a stony near-Earth asteroid, the most probable speed in the atmosphere is between 11 and 20 km/s and considering the small dimension $-$ below the cohesionless spin-barrier limit of about 150 m $-$, this leads us to hypothesize that most likely was a monolithic block. Monolithic blocks with dimensions up to several tens of meters are confirmed by the space exploration of asteroids, such as Ryugu \citep{Michikami2019} and Bennu \citep{Jawin2019}. \\
To estimate the dynamic pressures a fragment can encounter, we adopt the following qualitative model: an asteroid fall into the atmosphere as a rigid body losing mass by ablation until fragmentation takes place (when the dynamic pressure exceeds the effective strength): it follows a pancake phase until the diameter becomes sometimes larger than the original one, there is the airburst, and then start a fall phase for a fragment about one meter in diameter. The fragment is assumed to become an independent body at the end of the pancake phase, immediately after the airburst. We will assume that the initial speed of the fragment is the final speed of the pancake phase and that the fall direction remains the same as the main body. Finally, the strength of the fragment will be greater than the main body value according to Weibull's law (with $\alpha\approx 0.2$), although we set an upper limit of 100 MPa on the fragment strength.\\
In experiments with groups of spheres of identical diameter coupled together, it was found that the decoupling is complete when the diameter of the group of spheres that move away is about 2.28 times the original one \citep{Whalen2021} therefore, the ratio between the initial diameter and the final diameter of the pancake phase must be at least greater than this value. However, the further fragmentation that occurs in the pancake phase can delay the decoupling between the fragments so that a 5-10 times aspect ratio between the final diameter of the pancake phase and the original diameter (the so-called ``pancake factor'') is able to correctly describe the airburst altitude \citep{Chyba1993, Collins2005}. It should be noted that high pancake factor values, although empirically providing correct values for the airburst, are not considered physically realistic because - if interpreted literally - the pancake structure would become very thin compared to the diameter \citep{Collins2017}. As a reference value, we will assume 7.5, but in Table~\ref{tab:Tunguska_dynamic_pressure} results will also be reported for the extreme values, with pancake factor equal to 5 and 10.\\
For our purposes, we can use Eq.~(\ref{eq:motion_non_inertial}) for the atmospheric trajectory, with Eq.~(\ref{eq:mass_loss_ablation}) for the mass loss; Eq.~(\ref{eq:strength}) for the fragmentation condition; Eq.~(\ref{eq:dispersal_speed}) and Eq.~(\ref{eq:pancake_diameter}) for lateral expansion speed and increasing diameter in the pancake phase (immediately after the first fragmentation), and Eq.~(\ref{eq:scale_strength}) for the strength of the fragments after they become independent bodies. For the asteroid, we have adopted an average density of $3290 ~\textrm{kg}/\textrm{m}^3$, while for the fragment, we have adopted a fixed size of 1.4 m in diameter with a mass of 5000 kg, potentially able to resist the ablation phase during the airburst due to Eq.~(\ref{eq:ablation_time}). \\
Solving these equations numerically with Runge-Kutta 4th/5th order solver, we have a mathematical model for the fall of a stony asteroid into the atmosphere with pancake phase, airburst and big fragment fall. The software of our TCB fall model is available for download on GitHub\footnote{\url{ https://github.com/AlbinoCarbo/Tunguska_fall_model.git}}, the default settings file contains the values to obtain Fig.~\ref{fig:TE_model_15}. In this generic model we use the US Standard Atmosphere 1976 up to 86 km height\footnote{\url{https://www.pdas.com/atmos.html}}.

\noindent As a first test, we used our model to reproduce both the Chelyabinsk fall and the dynamics of the F1 fragment, assuming an initial mean strength of 4 MPa and a kinetic energy of 0.5 Mt for the main body. The initial conditions for trajectory and speed are given in Table~\ref{tab:Chelyabinsk_table}, instead of 97.1 km as $H_{start}$, we use 85 km in order to stay within the adopted atmospheric profile. This slightly lower value for the initial height has no effect on the final results shown in Fig.~\ref{fig:CE_model} because the speed loss in the initial stages is negligible. So the starting parameters are the following: initial height 85 km, speed 19.2 km/s, mean density $3290 ~\textrm{kg}/\textrm{m}^3$, diameter 20 m, inclination $18.3^\circ$ and main body mean strength 4 MPa. According to the computations, the fragmentation starts at 29.9 km, while the airburst occurs at a height of about 26.5 km. The pancake phase has a duration of 0.6 s, then our macroscopic fragment of 1.4 m initial diameter with strength from Weibull's law of 19 MPa and a speed of about 17 km/s continues the fall towards the ground. The fragment speed decreases rapidly, and the maximum dynamic pressure is 9 MPa at a height of about 20 km, so the fragment does not fragment further. There is still a slight mass loss by ablation that, in our model, ceases when the atmospheric speed reaches 3 km/s. At this point, after about 5.5 s from the end of the pancake phase, the dark flight phase begins at an altitude of about 12 km. After about 19 km of horizontal range, the meteoroid touches the ground with a diameter of about 0.8 meters at a speed of about 200 m/s. If we go back to Section~\ref{sec:CE}, we see that the numerical data of our model agree, with fair approximation, with the observed one. By doing a fine-tuning operation, the airburst height rises to 29.7 km if an average strength of 2 MPa is assumed, but the ``picture'' after the airburst remains essentially the same.

\begin{figure}[ht!]
\centering
\hspace*{-1.6cm}\includegraphics[width=1.2\hsize]{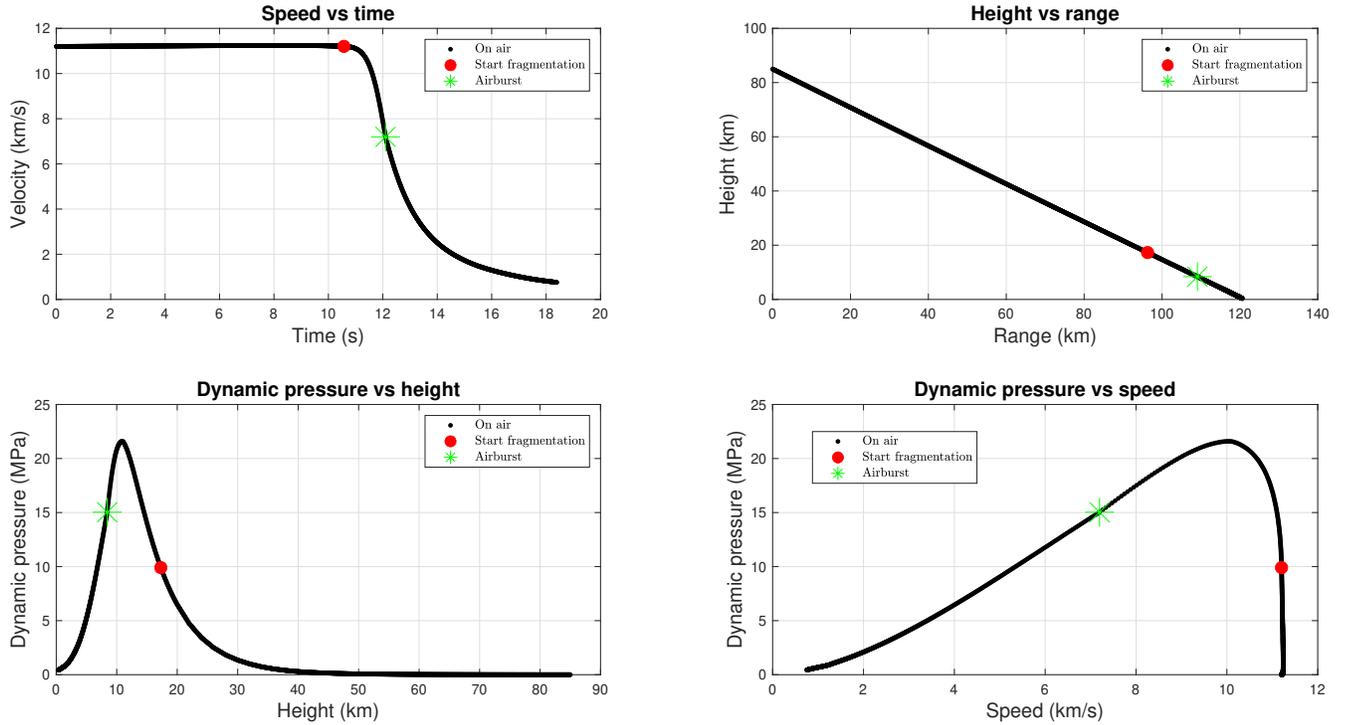}
\caption{The Tunguska model with initial kinetic energy 15 Mt, height 85 km, speed 11.2 km/s, $\Gamma \approx 0.58$, mean density $\rho_m = 3290 ~\textrm{kg}/\textrm{m}^3$, diameter 83 m, inclination $35^\circ$, pancake factor 7.5 and TCB's mean strength 10 MPa. Duration of the pancake phase 1.5 s, and airburst at 8.4 km. Fragment of 1.4 m diameter with a strength of 100 MPa, final diameter of 1.3 m and maximum dynamic pressure 21.5 MPa.}
\label{fig:TE_model_11}
\end{figure}

\noindent We proceeded similarly with the TE, but in this case, the mean strength is chosen in such a way to have an airburst in the range 8-9 km of altitude with an initial kinetic energy of 15 Mt, see Fig.~\ref{fig:TE_model_15} and Fig.~\ref{fig:TE_model_11}. The starting parameter of the first scenario are the following: initial height 85 km, speed 15 km/s, mean density $3290 ~\textrm{kg}/\textrm{m}^3$, diameter 69 m, inclination $35^\circ$, TCB's strength 25 MPa. According to the computations, the airburst occurs at a height of about 8.3 km, and the pancake phase (from the fragmentation to the airburst) has a duration of about 0.9 s. The maximum dynamic pressure is 40.4 MPa, reached during the pancake phase. Assuming that Weibull's law holds exactly, we can estimate the size of the largest fragment $D_{frmax}$ that will no longer break during the pancake phase, given by:

\begin{equation}
D_{frmax}=D_{main} \left( \frac{S_{main}}{P_{max}} \right)^{1/3 \alpha}
\label{eq:largest_fragment}
\end{equation}

\noindent In Eq.~(\ref{eq:largest_fragment}), derived from Eq.~(\ref{eq:scale_strength}), the maximum dynamic pressure $P_{max}$ replaces the strength $S_{fr}$ of the fragment. In this case we have $D_{main} \approx 69$ m, $S_{main} \approx 25$ MPa, $P_{max}\approx 40.4$ MPa and $\alpha\approx 0.2$, so $D_{frmax} \approx 30$ m. This cannot be true, otherwise there would be an impact crater.\\
As expected, there is a noticeable decrease in speed in the pancake phase, from 15 to 8.1 km/s. Our hypothetical fragment of 1.4 m with a maximum strength of 100 MPa can resist and continue the fall towards the ground with a residual speed of about 8.1 km/s and a slight mass loss by ablation. At an altitude of about 4 km with a speed of 3 km/s, the dark flight phase begins, and after about 7 km of horizontal range, the meteoroid touches the ground with a diameter of about 1.3 meters at a speed of about 0.8 km/s. \\
It is interesting to compare the final dimensions of the fragment in the Chelyabinsk case and the one in Tunguska: the Chelyabinsk fragment has smaller dimensions than the Tunguska, 0.8 m instead of 1.3 m. This is mainly due to the fragment's different initial speeds after this phase because the ablation increases with the cube of the velocity (see Eq.~(\ref{eq:mass_loss_ablation})). In the Chelyabinsk case, the starting speed of the fragment after the airburst is 17 km/s, while in the Tunguska case, it is only 8.1 km/s. This difference in initial speeds after the airburst, when the fragments are supposed to come out from the pancake phase, causes less ablation to the Tunguska fragment, so they arrive at the ground with a greater mass than the Chelyabinsk fragment.\\
The value for the impact speed is interesting because it is close to the estimated speed of about 0.5 km/s made by \cite{Anfinogenov2014} for the John's Stone by imposing that all the kinetic energy in the impact has done the work of excavation in the permafrost. Reversing Anfinogenov's formula, it can be estimated that a fragment of 1.3 m in diameter having a mass of about $m_{fr}\approx 3785$ kg hitting the permafrost with a strength $\sigma_{pf}\approx 3$ MPa at about $v_i\approx 0.8$ km/s would excavate a volume $B$ given by:

\begin{equation}
B \approx \frac{m_{fr}{v_i}^2}{2\sigma_{pf}} \approx 400 ~\textrm{m}^3
\label{eq:permafrost_volume}
\end{equation} 

\begin{figure}[ht!]
\centering
\hspace*{-1.6cm}\includegraphics[width=0.8\hsize]{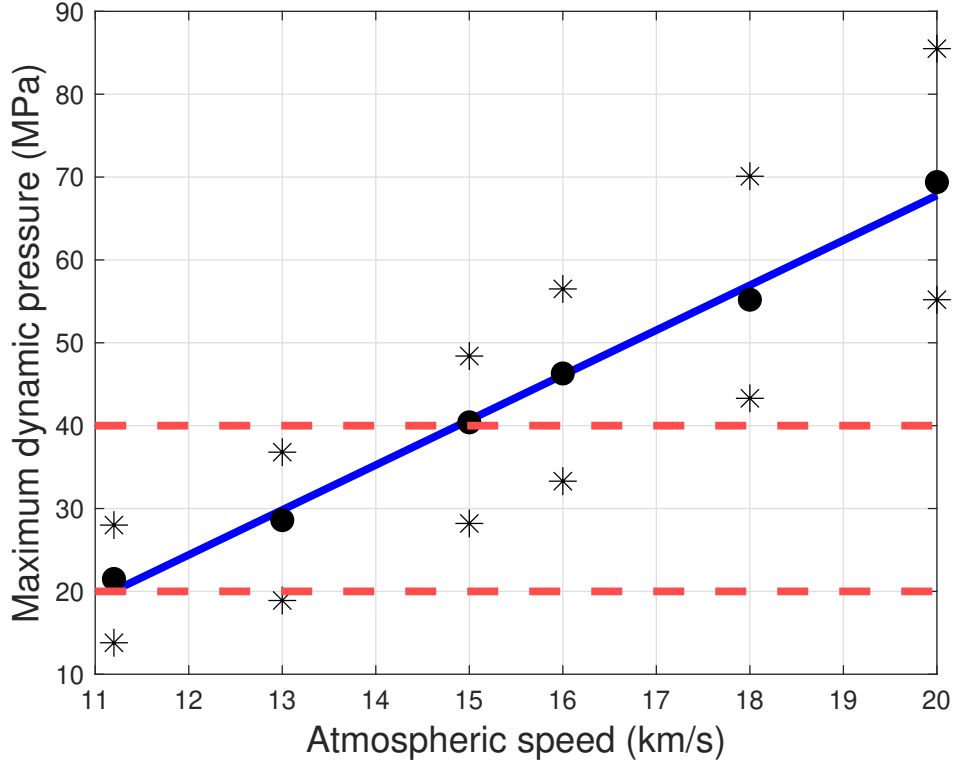}
\caption{The maximum dynamic pressure as a function of atmospheric entry speed for a TCB with kinetic energy of 15 Mt, trajectory inclination of $35^\circ$, pancake factor 7.5 and airburst height in the range 8-9 km. The two horizontal dotted lines indicate the lower and upper limits of the Carancas's mean strength, while the asterisks indicate the maximum dynamic pressure for models with pancake factors 5 (up) and 10 (down).}
\label{fig:Maximum_dynamic_pressure}
\end{figure}

This volume for the excavated permafrost is about 350 times the volume of the meteoroid, so it is reasonable to expect that the fragments became embedded in the permafrost and quickly swallowed up by the mud.\\
With an atmospheric entry speed of 11.2 km/s, the TCB's diameter is about 83 m, and the mean strength must be 10 MPa to have an airburst at about 8.4 km height. The maximum dynamic pressure is 21.5 MPa, achieved during the pancake phase (duration of 1.5 s), while that of the fragment is still 100 MPa (see Fig.~\ref{fig:TE_model_11}). The macroscopic fragment reaches the ground with a speed of about 0.7 km/s, a diameter of about 1.3 m and a mass of 3785 kg. In this case, the volume $B$ excavated in the permafrost and given by Eq.~(\ref{eq:permafrost_volume}) is about $300~\textrm{m}^3$, 260 times the volume of the fragment. \\
Finally, we have computed the model for a speed of 20 km/s and a diameter of 57 m (kinetic energy of 15 Mt). In this case, to have an airburst at about 8.8 km, the TCB's mean strength must rise to about 50 MPa. So the maximum dynamic pressure is 69.4 MPa, and our 1.4 m fragments, with a maximum strength of 100 MPa, can reach the ground with a speed of about 0.5 km/s and a diameter of about 1.3 m. The maximum dynamic pressures vs some values of atmospheric entry speed are given in Table~\ref{tab:Tunguska_dynamic_pressure} and plotted in Fig.~\ref{fig:Maximum_dynamic_pressure}.\\

\begin{table}
\centering
\caption{Tunguska fall models computed for a TCB with an initial kinetic energy of 15 Mt, mean density of $3290 ~\textrm{kg}/\textrm{m}^3$, trajectory inclination of $35^\circ$, airburst height in the range 8-9 km and pancake factor 5.0 (up), 7.5 (middle), 10 (bottom). The first column on the left is the starting atmospheric entry speed of the TCB, the second is the average body strength to get the airburst in the right altitude range, the third is the maximum dynamic pressure that is reached during the pancake phase, and finally, the last two columns are the fragment's strength based on Weibull's law (but with a superior limit of 100 MPa) and the fragment's starting speed after the airburst. In almost any scenario, the mean strength of the fragment reaches the maximum allowed value and the maximum dynamic pressure is at most twice the maximum strength of the Carancas fall.}
\label{tab:Tunguska_dynamic_pressure}

\begin{tabular}{lcccc}
\hline
$V$ (km/s) & $S_{main}$ (MPa) & $Max P_{dyn}$ (MPa) & $S_{fr}$ (MPa) & $v_{fr}$ (km/s)\\
\hline
11.2         & 18               & 28.0                &  100           & 10.0\\
13.0         & 25               & 36.8                &  100           & 11.5\\
15.0         & 35               & 48.4                &  100           & 13.1\\
16.0         & 43               & 56.5                &  100           & 13.9\\
18.0         & 55               & 70.1                &  100           & 15.5\\
20.0         & 70               & 85.5                &  100           & 16.9\\
\hline
\hline
11.2         & 10               & 21.5                &  100           & 7.2\\
13.0         & 15               & 28.6                &  100           & 7.9\\
15.0         & 25               & 40.4                &  100           & 8.1\\
16.0         & 30               & 46.3                &  100           & 8.0\\
18.0         & 37               & 55.2                &  100           & 7.5\\
20.0         & 50               & 69.4                &  100           & 5.0\\
\hline	
\hline
11.2         & 3                & 13.8                &  34            & 3.5\\
13.0         & 6                & 18.9                &  65            & 3.0\\
15.0         & 13               & 28.2                &  100           & 1.8\\
16.0         & 17               & 33.3                &  100           & 1.2\\
18.0         & 25               & 43.3                &  100           & 0.3\\
20.0         & 35               & 55.2                &  100           & 0.1\\
\hline
\end{tabular}
\end{table}

To conclude this section, in our working hypothesis, the TCB was a monolithic stony asteroid with an atmospheric entry speed in the range of 11-20 km/s and a mean strength in the range of 3-70 MPa (to have the airburst in the 8-9 km range), up to 17 times higher than the average strength of Chelyabinsk which we estimated in about 4 MPa. A macroscopic fragment, with a diameter of about 1 m and strength in the range of 14-85 MPa (Fig.~\ref{fig:Maximum_dynamic_pressure} and Table~\ref{tab:Tunguska_dynamic_pressure}), could have sufficient dimension and strength to avoid vaporization and further fragmentation. A strength in this range is physically compatible with the mechanical strength measured in laboratory on stony meteorites: from 6 to 200 MPa for compressive strength and from 2 to 62 MPa for tensile strength \citep{Svetsov1995}. The fall of Carancas, which took place in Peru on September 15, 2007, is emblematic: in this case, there was no fragmentation, and the stony meteoroid arrived intact on the ground, generating a crater 13 m in diameter near Lake Titicaca \citep{Borovicka2008}. The estimated starting speed for this monolithic body, with dimensions of 0.9-1.7 m, is under 23 km/s and the dynamic strength ranges from 20 to 40 MPa, values much higher than 0.4-12 MPa valid for small meteoroids. A TCB's fragment with strength about 2 times the maximum value estimated for the Carancas fall appears to be sufficient to survive Tunguska's airburst even in the worst-case scenario, and the probability of survival increases as the TCB's atmospheric entry speed decreases. The macroscopic fragments reach the ground at a speed in the range of 0.8-0.5 km/s and get stuck in the permafrost. This scenario would explain the initial eyewitness accounts and the subsequent disappearance of the meteorites due to incorporation into the permafrost.

\begin{table}
\centering
\caption{The nominal coordinates of the strewn fields center as a function of the starting $m_{fr}$ with a trajectory inclination of $35^\circ \pm 5^\circ$, azimuth $110^\circ \pm 10^\circ$, speed $V=10 \pm 3$ km/s, height $h=8.5 \pm 1$ km and $\Gamma=0.775$. The starting diameter, $D_0$ is computed assuming a mean density equal to Chelyabinsk, $\rho = 3290~\textrm{kg}/\textrm{m}^3$. The range of masses considered goes from the Chelyabinsk F1 fragment to Carancas.}
\label{tab:Tunguska_fragments}

\begin{tabular}{lccccc}
\hline
$m_{fr}~(\textrm{kg})$ & $m_{fr}/A_0~(\textrm{kg}/\textrm{m}^2)$ & $D_0$ (m)  & Lat. N ($^\circ$) & Long. E ($^\circ$)\\
\hline
6000     & 3300         & 1.5        &$60.921 \pm 0.017$ & $101.696 \pm 0.026$\\
5000     & 3120         & 1.4        &$60.921 \pm 0.017$ & $101.697 \pm 0.027$\\
4000     & 2900         & 1.3        &$60.921 \pm 0.017$ & $101.698 \pm 0.026$\\
590      & 1530         & 0.7        &$60.916 \pm 0.015$ & $101.724 \pm 0.022$\\
\hline	
\end{tabular}
\end{table}

\section{Tunguska's strewn field}
\label{sec:TE_strewn_field}
\begin{figure}[ht!]
\centering
\hspace*{-0.8cm}\includegraphics[width=0.8\hsize]{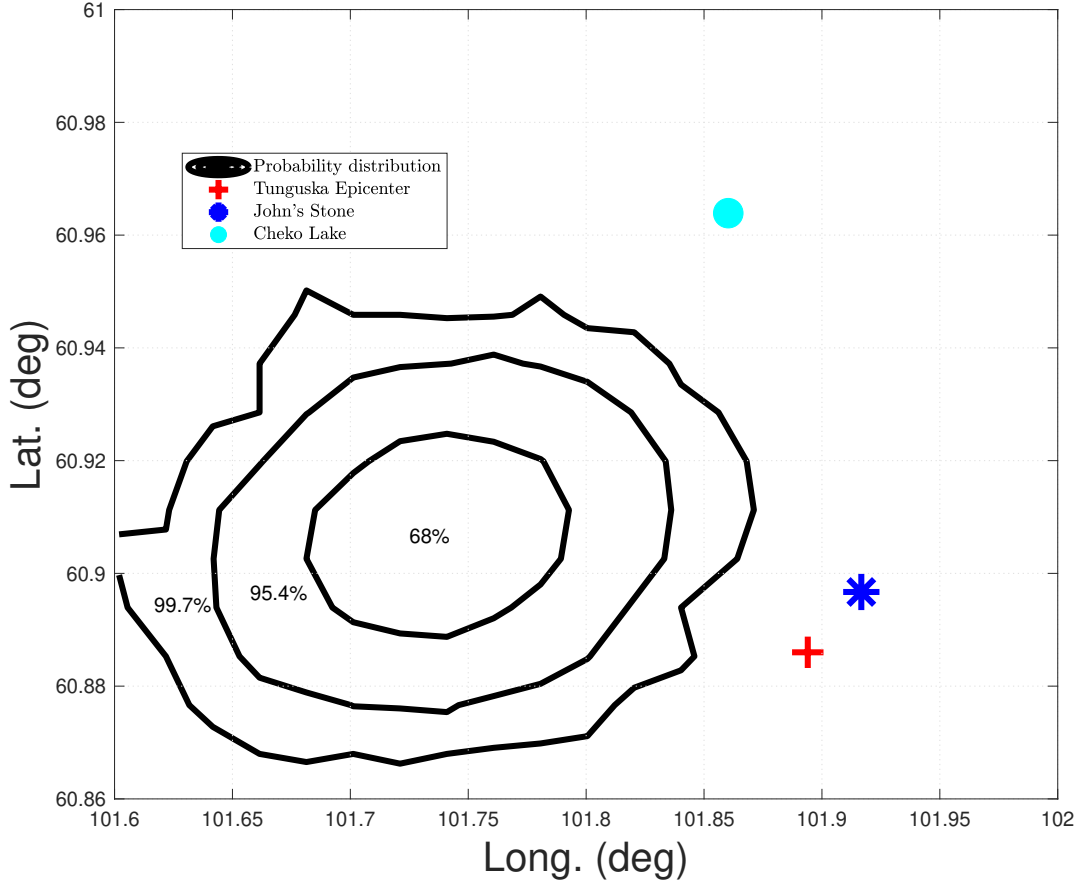}
\caption{The probability distribution of the strewn field for the TE computed for fragments with starting mass $m_{fr} = 590, 4000, 5000 ~\textrm{and}~6000~\textrm{kg}$ as in Table~\ref{tab:Tunguska_fragments}. For each $m_{fr}$ value, 5000 Monte Carlo scenarios with standard normal distribution were computed with inclination $35^\circ \pm 5^\circ$, azimuth $110^\circ \pm 10^\circ$, speed $V=10 \pm 3$ km/s, height $h=8.5 \pm 1$ km, latitude $60.886^\circ \pm 0.002^\circ$, longitude $101.894^\circ \pm 0.002^\circ$ and $\Gamma=0.775$. The internal curve encloses an area with a fall probability of 68\% (1 sigma), the intermediate one of 95.4\% (2 sigma), while the outermost one represents the zone with a fall probability of 99.7\% (3 sigma). Cheko Lake is about 3.5 km from the edge of the strewn field.}
\label{fig:TE_strewn_field2}
\end{figure}

As in the case of Chelyabinsk, to compute the path followed by the possible macroscopic fragments that survived the airburst is necessary to know the height, azimuth, inclination, latitude and longitude of the starting point (i.e. the epicenter) and the fragment's starting speed. The geometrical parameters, with their uncertainties, are taken from Table~\ref{tab:Tunguska_table}. About the fragment's starting speed, we take $V=10 \pm 3$ km/s in order to cover most of the possible speed values of Table~\ref{tab:Tunguska_dynamic_pressure}. \\
The mathematical model used for strewn field computation is from Eq.~(\ref{eq:motion_non_inertial}) with $\Gamma = 0.775$ and $\rho_m = 3290 ~\textrm{kg}/\textrm{m}^3$, as before. To this equation must be added Eq.~(\ref{eq:mass_loss_ablation}) for mass ablation because, as we saw in the previous section, after the airburst, the fragments have a speed mostly higher than 3 km/s, the ablation limit. The computations were made for four starting values of the fragment's mass: $m_{fr} = 6000, 5000, 4000, 590~\textrm{kg}$: these values go from the Chelyabinsk F1 fragment to Carancas and include the mass of our reference fragment (5000 kg) considered in Section~\ref{sec:maximum dynamic pressure}.\\
As an atmospheric profile, we chose the NRLMSISE-00 model \citep{Picone2002}, computed for the coordinates and the date of the atmospheric explosion. Considering that we aim to trace the largest fragments shown in Table~\ref{tab:Tunguska_fragments}, the exact wind profile is not very important, and this is a fortunate circumstance because there are no atmospheric profiles of wind speed and direction for the TE. To compute the uncertainty of the strewn field, we used a Monte Carlo approach with standard normal distribution, generating 5000 different scenarios for each fragment and looking at the geographic distribution of the possible fall zones (see Fig.~\ref{fig:TE_strewn_field2}). \\
About TE there are important differences from the low inclination of the trajectory, estimated by the eyewitnesses, and that obtained from the analysis of the devastation area. As we have already mentioned at the beginning of Section~\ref{sec:TE}, in this last case, the best estimate is for angles equal to $35^\circ$, about double the $15^\circ$ eyewitness value, see Table~\ref{tab:Tunguska_table}. Considering that in Section~\ref{sec:maximum dynamic pressure}, the models were computed for a high inclination angle, we will consider only a high angle for the strewn field.\\
So, the parameter varied in Monte Carlo routine are the following: height ($8.5 \pm 1$ km), azimuth ($110^\circ \pm 10^\circ$), inclination ($35^\circ \pm 5^\circ$), latitude of the starting point ($60.886^\circ \pm 0.002^\circ$ N), longitude of the starting point ($101.894^\circ \pm 0.002^\circ$ E) and the fragment's starting speed ($10 \pm 3$ km/s).

\noindent The center of the mean strewn field (Lat. $60.919^\circ \pm 0.002^\circ$; Long. $101.70^\circ \pm 0.01^\circ$) is about 11 km WNW away from the epicenter, see Table~\ref{tab:Tunguska_fragments}, and forms an ellipse quite large: at 3 sigma level the major axes are about 16 and 11 km, for a total area of about $\pi \cdot 8 \cdot 5.5 \approx 140~\textrm{km}^2$.

\begin{figure}[ht!]
\centering
\hspace*{-0.8cm}\includegraphics[width=0.8\hsize]{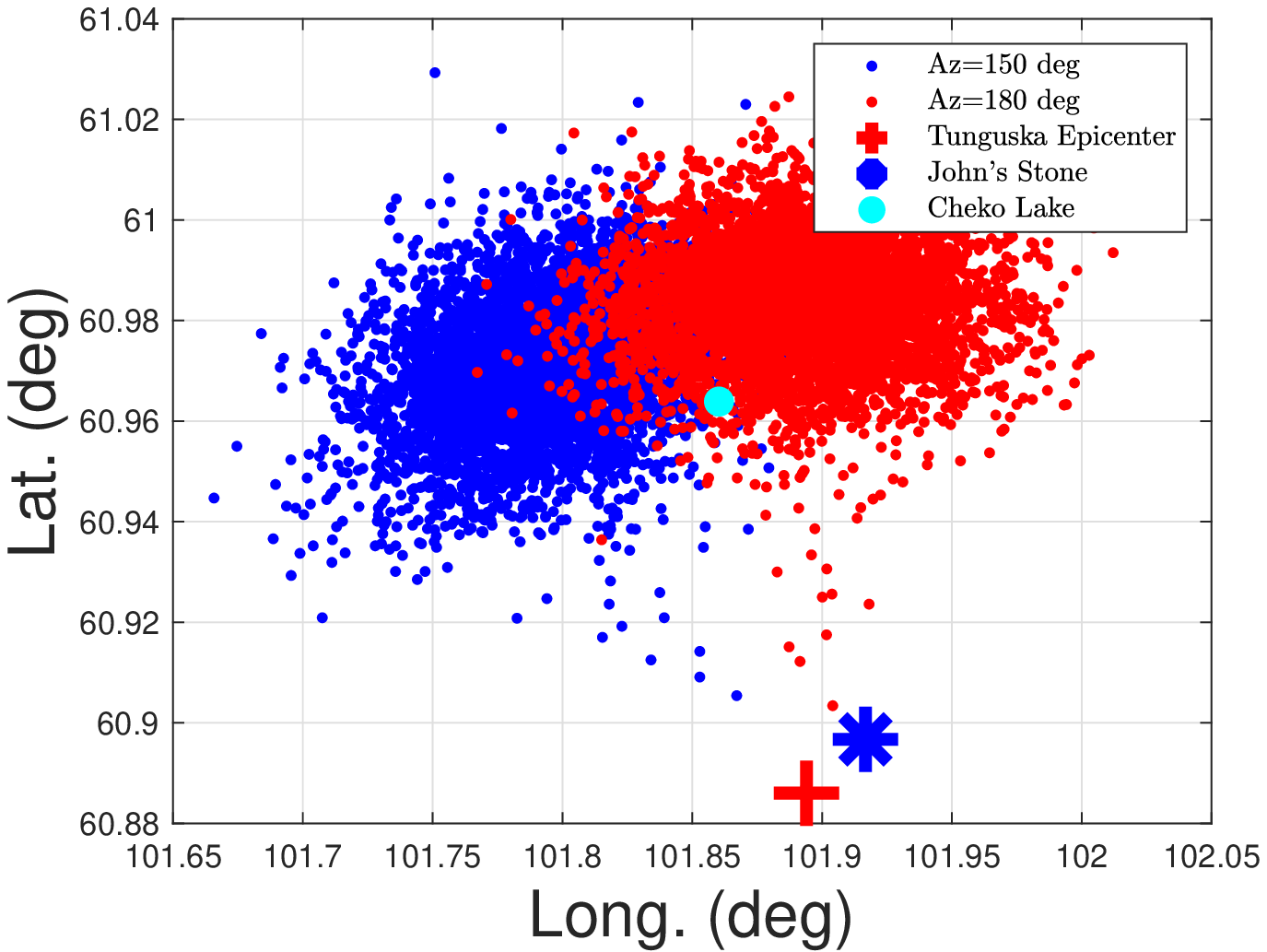}
\caption{The strewn fields for Lake Cheko for fragments with $m_{fr} = 5000~\textrm{kg}$: 5000 Monte Carlo scenarios with standard normal distribution were computed with inclination $35^\circ \pm 5^\circ$, azimuth $(150^\circ-180^\circ) \pm 10^\circ$, speed $V=10 \pm 3$ km/s, height $h=8.5 \pm 1$ km and $\Gamma=0.775$.}
\label{fig:Cheko_strewn_field}
\end{figure}

\subsection{About the Lake Cheko origin}
These results about a possible strewn field for TE allow us to make some comments about the Cheko Lake, a small lake elliptical in shape with axes equal to $500 \times 350$ m elongated in NW-SE direction, located about 9 km in a direction with an azimuth of $348^\circ$ from the epicenter with coordinates Long. $101.8603^\circ$ E, Lat. $60.964^\circ$ N. Some authors consider Cheko Lake as a possible impact crater of a fragment of the TCB with a diameter of about 1 m \citep{Gasperini2007, Gasperini2008, Gasperini2009, Gasperini2014}. The most intriguing clues about this lake are the conical shape, the density/velocity discontinuity in the center of the lake about 10 m below the bottom (the so-called reflector-T) and the trees near the shores of the lake that show vigorous growth only after 1908 until the present day, suggesting that they are born when there was no lake \citep{Gasperini2014, Fantucci2015}. The scientific debate about Lake Cheko's origin is still open, and several authors do not consider it as an impact structure: there is no sure evidence that the lake is only 100 years old, and it is strange that a single impact structure formed during the TE because crater fields are usually found, furthermore the edge of the lake is not as prominent as one would expect from an impact crater \citep{collins2008}.\\
On the basis of our findings, Cheko Lake is outside of about 3.5 km from the more likely strewn field at 3 sigma level, see Fig.~\ref{fig:TE_strewn_field2}). Making computations of the strewn field by radically changing the trajectory's azimuth, with $i=35^\circ\pm 5^\circ$, $V=10 \pm 3$ km/s, $h=8.5 \pm 1$ km, $\Gamma=0.775$ and $m_{fr} \approx 5000~\textrm{kg}$, we find that the lake could be an impact crater if the azimuth is between 150 and 180 degrees, see 
Fig.~\ref{fig:Cheko_strewn_field}. These values agree with the fact that to join the epicenter with the lake, the incoming trajectory would have an azimuth of $348^\circ - 180^\circ \approx 168^\circ$. However, these values differ from the most common azimuth value adopted here ($110^\circ \pm 10^\circ $). The value for the azimuth of the trajectory closest to $168^\circ$ is that of $137^\circ$ obtained by \cite{Krinov1949}, based on eyewitness accounts of the fireball. On the other hand, it is unlikely that the value of $110^\circ \pm 10^\circ $ be far from reality. The area of the trees fall resembles a triangle or a butterfly and has an axis of symmetry directed towards an azimuth of about $115^\circ $ \citep{artemieva2016}: with this condition, it is difficult to think that the trajectory of the TCB could have been directed towards the lake. A deviation of at least $40^\circ$ would be necessary, but this is almost impossible for a macroscopic fragment generated in the epicenter. In fact, from Eq.~(\ref{eq:dispersal_speed}), the deflection angle $\theta_{disp}$ of the speed vector following the pancake phase with lateral expansion, with $\rho_a\approx 0.5 ~\textrm{kg}/\textrm{m}^3$ and $\rho_m = 3290 ~\textrm{kg}/\textrm{m}^3$, is given by:

\begin{equation}
\theta_{disp} \approx \arctan \left( \sqrt{\frac{7}{2}\frac{\rho_a}{\rho_m}} \right) 
\label{eq:dispersal_angle}
\end{equation}

\noindent From Eq.~(\ref{eq:dispersal_angle}) we have $\theta_{disp}\approx \arctan \left(0.023\right) \approx 1.3^\circ$, more or less the same deviation value of the Chelyabinsk fragment F1. However, considering the geometry formed by the epicenter, Lake Cheko and arrival trajectory from $110^\circ$ azimuth, in order for a fragment formed in the 8.5 km airburst to impact the area of Lake Cheko to form a crater, a deviation of about $58^\circ$ from the direction of arrival is required, a value that is consistent with Fig.~\ref{fig:Cheko_strewn_field}.\\

\section{Conclusions}
\label{sec:end}
The purpose of this paper was to delimit a possible strewn field for Tunguska where to go in search of macroscopic fragments large enough to survive the high temperatures and pressure during the airburst phase. In the case of the Chelyabinsk fall, a stony meteorite with a diameter of this order of magnitude was recovered (the so-called F1 fragment), and the same thing happened in the fall of Carancas. \\
Using the Chelyabinsk event as a reference, we have tested a model based on Earth's gravity and atmospheric drag, able to give, with a good approximation, the dark flight and the fall zone of the fragment F1. From the strength of the meteorites recovered from CE, using Weibull's law with modulus $\alpha\approx 0.2$, a strength of the original body $S\approx 4$ MPa was estimated. We have adopted the same $\alpha$ value for the Tunguska case. After a review of the models used for the description of the Tunguska fall and having verified that fragments with a diameter of the order of 1 m can survive the airburst phase if they have enough strength, a falling model has been developed which includes mass ablation, pancake phase with lateral expansion and Weibull's law for fragments strength. \\
The results, for a TCB with a kinetic energy of 15 Mt, atmospheric entry speed in the range 11-20 km/s and a mean strength in the range 3-70 MPa to have the airburst in the 8-9 km range, tell us that a macroscopic fragment with a mean strength in the range 14-85 MPa and a starting mass of 5000 kg (diameter of about 1.4 m), would be sufficient to reach the ground. The maximum strength of the fragment is about 2 times the maximum strength estimated for the fall of Carancas, a value that is physically possible. The fragment's arrival speed on the ground is in the range of 0.8-0.5 km/s, high enough to get stuck in the permafrost and be quickly lost.\\
The mean strewn field center for high trajectory inclination ($35^\circ \pm 5^\circ$) and for masses that goes from the Chelyabinsk F1 fragment to Carancas, is about 11 km WNW with an extension of about $140~\textrm{km}^2$ km. If some macroscopic meteorites were produced, this is the most likely region to look for. Unfortunately, the time elapsed from the fall of the TCB to the first Kulik expedition was 19 years, enough time for any little craters and meteorites to be buried by mud and vegetation, which is even more true today. \\
Lake Cheko is about 3.5 km outside the strewn fields at 3 sigma level and, based on our results, it is unlikely that it could be a real impact crater unless the trajectory of the TCB had an azimuth between $150^\circ$ and $180^\circ$. However, this value contrasts with the azimuth of the axis of symmetry of the area of the trees fall, which indicates the direction of the most probable arrival trajectory.

\section*{Acknowledgements}
The authors want to thank very much an anonymous referee and Gareth Collins, whose suggestions make this paper much better than the original.

\section*{Data Availability}
The data underlying this paper will be shared on reasonable request to the corresponding author.

%-------------------------------------------------------------------

\end{document}